\PassOptionsToPackage{hyphens}{url}
\documentclass[10pt,twocolumn,letterpaper]{article}

\usepackage[T1]{fontenc}
\usepackage{subcaption}
\usepackage{newfloat}
\usepackage{listings}
\usepackage{multirow}
\usepackage{amssymb}
\usepackage{bm}
\usepackage{algorithm}
\usepackage{algorithmic}
\usepackage{booktabs}
\usepackage{enumerate}
\usepackage{makecell} 
\usepackage{graphicx}
\usepackage{mathtools}
\usepackage{url}
\usepackage{amsmath}
\usepackage{amssymb}
\usepackage{xspace}
\usepackage{verbatim}

\usepackage[pagenumbers]{cvpr} 

\definecolor{cvprblue}{rgb}{0.21,0.49,0.74}
\usepackage[pagebackref,breaklinks,colorlinks,allcolors=cvprblue]{hyperref}
\makeatletter
\newtheorem{definition}{Definition}

\makeatother
\def\paperID{5813} 
\def\confName{ICCV}
\def\confYear{2025}

\title{SDF-TopoNet: A Two-Stage Framework for Tubular Structure Segmentation via SDF Pre-training and Topology-Aware Fine-Tuning}

\author{
    Siyi Wu\textsuperscript{1}\thanks{Equal contribution.} \quad
    Leyi Zhao\textsuperscript{2}\thanks{Equal contribution.} \quad
    Haotian Ma\textsuperscript{1} \quad
    Xinyuan Song\textsuperscript{3}\thanks{Corresponding author.} \\
    \textsuperscript{1} University of Texas at Arlington \quad
    \textsuperscript{2} Indiana University \quad
    \textsuperscript{3} Emory University \\
    \texttt\small \{sxw8121, 
hxm3470\}@mavs.uta.edu,\{leyizhao\}@iu.edu,\{xsong30\}@emory.edu
}

\date{}

\begin{document}
\maketitle

\begin{abstract} 
    Accurate segmentation of tubular and curvilinear structures, such as blood vessels, neurons, and road networks, is crucial in various applications. A key challenge is ensuring topological correctness while maintaining computational efficiency. Existing approaches often employ topological loss functions based on persistent homology, such as Betti error, to enforce structural consistency. However, these methods suffer from high computational costs and are insensitive to pixel-level accuracy, often requiring additional loss terms like Dice or MSE to compensate. To address these limitations, we propose \textbf{SDF-TopoNet}, an improved topology-aware segmentation framework that enhances both segmentation accuracy and training efficiency. Our approach introduces a novel two-stage training strategy. In the pre-training phase, we utilize the signed distance function (SDF) as an auxiliary learning target, allowing the model to encode topological information without directly relying on computationally expensive topological loss functions. In the fine-tuning phase, we incorporate a dynamic adapter alongside a refined topological loss to ensure topological correctness while mitigating overfitting and computational overhead. We evaluate our method on five benchmark datasets. Experimental results demonstrate that SDF-TopoNet outperforms existing methods in both topological accuracy and quantitative segmentation metrics, while significantly reducing training complexity. Our code is available at \href{https://github.com/siyiwu0330/SDF-TopoNet/}{SDF-TopoNet}. 
\end{abstract}

\section{Introduction}

\begin{figure}[!ht]
    \centering
    \resizebox{\columnwidth}{!}{
        \includegraphics{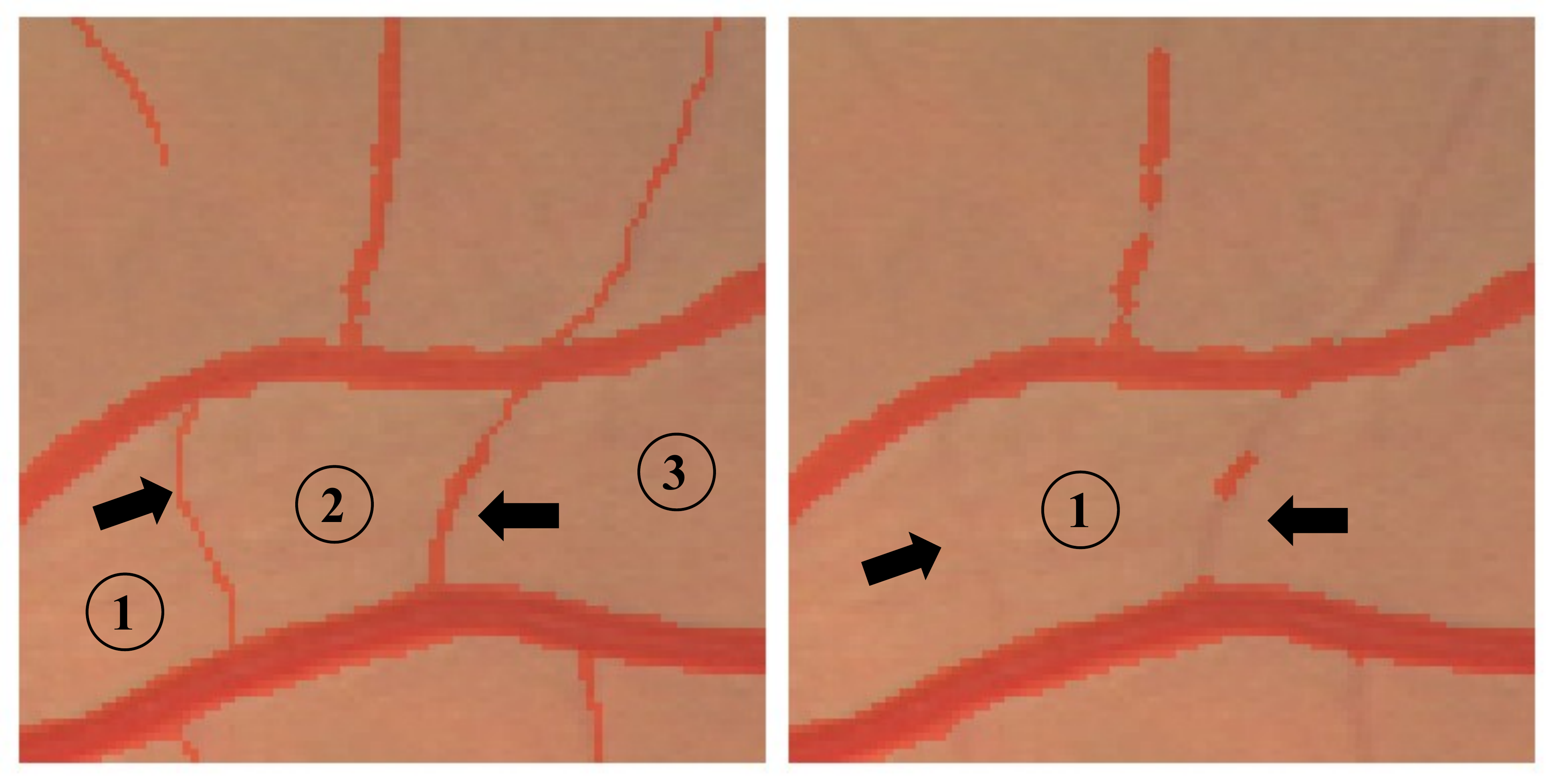}
    }
    \caption{An illustration of topological accuracy in retinal vessel segmentation. The left image represents the expected segmentation (ground truth), and the right image shows a possible prediction using a pixel-based loss function. }
    \label{fig:motivation}
\end{figure}

Recent advancements in deep learning-based segmentation have significantly improved pixel-wise accuracy \cite{long2015fully,chen2017deeplab,guo2022segnext,kirillov2023segment}. However, standard loss functions such as cross-entropy and Dice loss~\cite{li-etal-2020-dice} primarily focus on pixel-level supervision and fail to enforce global topological consistency. This limitation is particularly evident in the segmentation of thin and complex structures, where minor discontinuities can lead to topological errors, such as incorrect Betti numbers, resulting in missing or spurious connections. For example, even state-of-the-art models, such as the Segment Anything Model (SAM)~\cite{kirillov2023segment}, struggle with accurately segmenting topological structures like blood vessels and neuronal pathways \cite{wu2023medical}. Figure \ref{fig:motivation} illustrates how purely pixel-based losses fail to preserve topological continuity, causing segmentation gaps that disrupt connectivity. Specifically, for major vessels with distinct boundaries, the mask covers them well. However, because small vessels often have a lower likelihood during the prediction process, the mask may break in the middle. This breakage effectively changes the Betti number between the two major vessels from 3 to 1.

To address this issue, topological data analysis (TDA) has been integrated into segmentation models to maintain topological consistency. Methods leveraging Betti numbers, persistent homology, and cubical complexes have demonstrated improvements in capturing global structural properties. A common approach is incorporating topological loss functions that measure discrepancies between the predicted segmentation and the ground truth in terms of topological features (e.g. clDice \cite{shit2021cldice}, TopoLoss \cite{hu2019topology}, and Betti Matching \cite{stucki2023topologically}). While these methods have led to promising results, they introduce two significant challenges:  

\begin{enumerate}
    \item \textbf{High computational complexity:} Topological losses like Betti matching and persistence diagram-based losses require expensive operations that scale poorly with image size. As a result, prior works have often restricted training to small patches (e.g., $65\times 65$ for TopoLoss, and $48\times 48$ for Betti matching).
    \item \textbf{Insensitivity to local pixel-wise errors:} Topological losses focus on preserving global structural properties by minimizing topological errors such as Betti number discrepancies. However, they do not explicitly penalize small pixel-wise perturbations, making them insufficient for ensuring precise boundary adherence. As a result, they are often combined with conventional pixel-wise losses such as Dice or MSE to maintain segmentation accuracy.
\end{enumerate}

To overcome these limitations, we propose \textbf{SDF-TopoNet}, a novel two-stage training framework that enhances topological segmentation accuracy while reducing computational overhead. Instead of directly applying topological loss throughout training, we introduce a pre-training phase where the model learns to predict the signed distance function (SDF) of the target mask. Since SDF inherently encodes distance information relevant to persistent homology, this step allows the model to capture topological structure without explicitly computing topological loss, significantly reducing early-stage training complexity. In the fine-tuning phase, we then introduce a dynamic adapter and apply a refined topological loss to ensure the final segmentation preserves both pixel-level precision and topological correctness.

Our main contributions are summarized as follows:

\begin{itemize}

    \item We propose a two-stage training framework \textbf{SDF-TopoNet:} pre-training with MSE on SDF predictions followed by fine-tuning with a topological loss. This design reduces the computational burden associated with topological loss while improving overall segmentation quality.
    
    \item We propose the use of signed distance functions (SDFs) as an intermediate representation, enabling the model to learn rich boundary and topological information in a computationally efficient manner.
    
    \item We replace the conventional segmentation head with a dynamic adapter layer specifically designed to process SDF outputs. This layer adaptively learns the optimal transformation to convert the learned SDF into a binary segmentation mask during the tuning phase.
    
    \item Through extensive experiments on five public benchmark datasets demonstrate the effectiveness of our approach. SDF-TopoNet achieves superior topological accuracy while preserving pixel-level precision, all with reduced training costs.

\end{itemize}

\section{Related Work}

Segmenting tubular data, such as blood vessels \cite{zhang2019attention}, neurons \cite{li2019precise}, and roads \cite{zhou2018d}, has received growing attention in the field of computer vision. Traditional approaches often focus on optimizing pixel-level overlaps, which can result in disconnected or topologically incorrect predictions, especially in elongated or branching targets. To address these issues, several topology-aware methods have been introduced to preserve structural continuity. For instance, Mosinska et al. \cite{mosinska2018beyond} applied filtered responses in a pre-trained VGG19 to promote connectivity, and Shit et al. \cite{shit2021cldice} proposed clDice, which uses a differentiable skeleton extraction procedure to align predicted masks with ground truth skeletons. Arnavaz et al. \cite{arnavaz2021semi} further introduced a skeleton-based loss function to reinforce topological features.

Persistent homology has played a central role in many of these solutions. Approaches based on persistent homology \cite{carlsson2009topology,hensel2021survey} provide a multi-scale representation of topological features. \cite{hu2019topology} introduced the Wasserstein loss, which adapts the Wasserstein distance to improve image segmentation by matching dimension-1 points in persistence diagrams of the ground truth and the prediction. Beyond that loss, Clough et al. \cite{clough2020topological} discuss expected Betti numbers to reduce computational complexity, and Stucki et al. \cite{stucki2023topologically} propose Betti matching for more spatially meaningful correspondences. Other research draws on discrete Morse theory \cite{delgado2014skeletonization}, homotopy warping \cite{jain2010boundary, hu2021topoaware}, and marker-based losses \cite{wagner2011efficient} to maintain connectivity. Persistent homology also appears beyond segmentation, including crowd localization \cite{abousamra2021localization} and 3D cell shape reconstruction \cite{waibel2022capturing}.

Other studies use pixel-overlap measurements of topologically important structures to maintain continuity. For example, Hu et al. \cite{hu2019topology} utilize discrete Morse theory to isolate topologically critical regions within predictions and ground truth. Marker-based approaches have also been explored, such as Wang et al. \cite{wang2022ta}, who use a marker loss derived from Dice to enhance the connectivity of gland segmentations. Although these overlap-based methods are computationally efficient, they do not explicitly guarantee a spatial match of features. Additional strategies focus on embedding connectivity priors \cite{sasaki2017joint,wang2018post}, employing specialized filters \cite{mosinska2018beyond}, or incorporating template masks to enforce specific diffeomorphism types \cite{zhang2022topology}. Meanwhile, Cheng et al. \cite{cheng2021joint} propose an iterative feedback mechanism to model connectivity, and Oner et al. \cite{oner2020promoting} aim to improve topological accuracy by segmenting regions around curvilinear structures.

\section{Methodology}

Our method builds on concepts from topological data analysis (TDA) to enhance segmentation accuracy while maintaining computational efficiency. To ensure clarity, we provide an overview of the relevant mathematical foundations in Supp. Mat.~\cref{appendix:background}, including homology, Betti numbers, persistent homology, and the computation of persistence diagrams.

SDF-TopoNet is designed to improve segmentation accuracy by leveraging a two-stage training framework that integrates pixel-wise learning with topological constraints while mitigating computational overhead. Our approach consists of two key phases:

\begin{enumerate}
    \item \textbf{Pre-training Phase}: The model learns structural and boundary representations by predicting the signed distance function (SDF) of the target structures using an MSE loss. This step enables the network to encode topological information without explicitly relying on costly topological losses.
    \item \textbf{Fine-tuning Phase}: We replace the original segmentation head with a dynamic adapter. The model is then optimized using a weighted combination of topological loss and Dice loss, ensuring both topological and pixel-wise accuracy.
\end{enumerate}

\begin{figure}[!ht]
\centering
\resizebox{0.95\linewidth}{!}{
\includegraphics[width=1.0\linewidth]{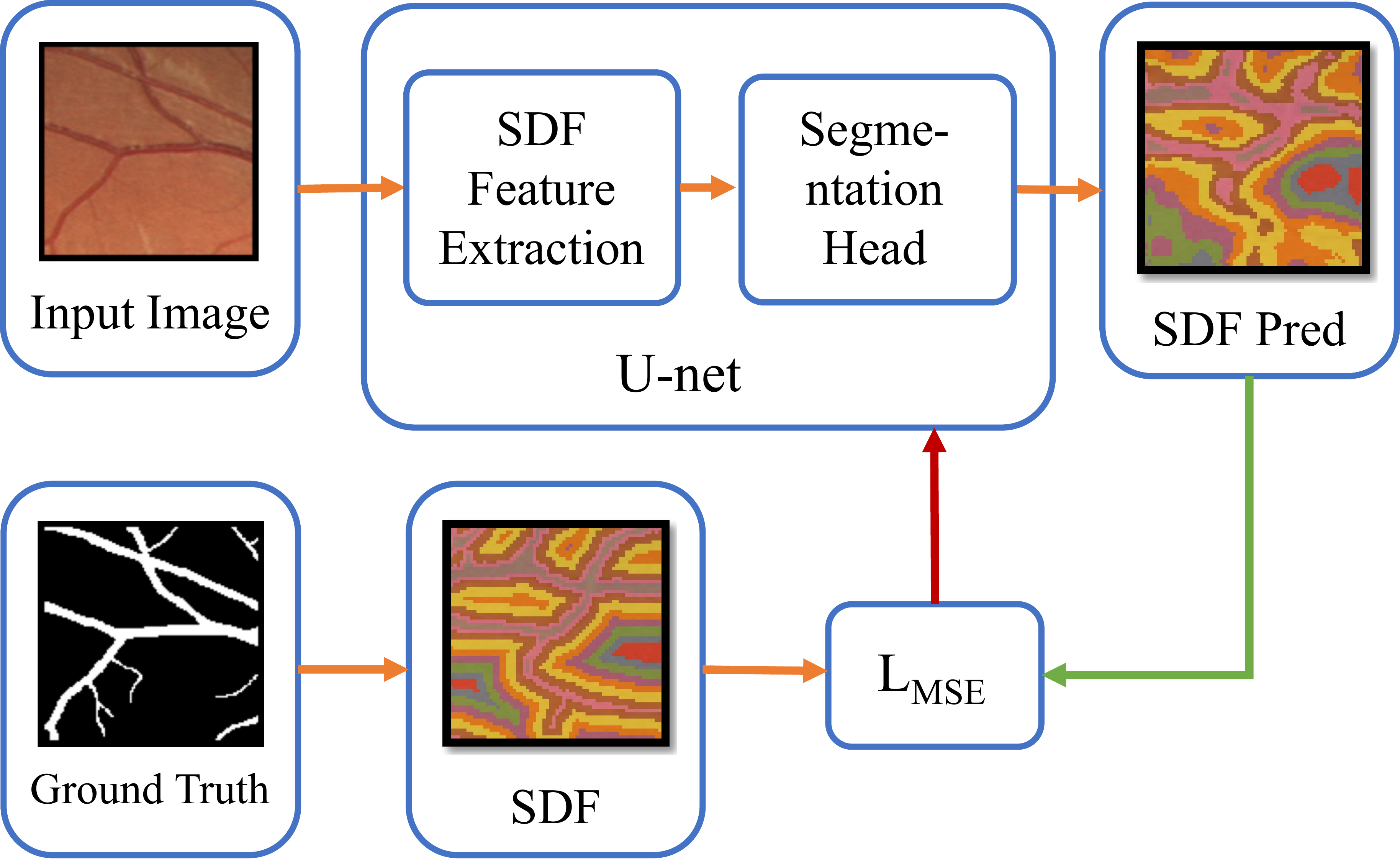}}
\caption{DF-TopoNet pre-training phase. The U-Net predicts the signed distance function (SDF) from the input image, optimizing with MSE loss to learn structural features.}
\label{fig:pre_train_pipeline}
\end{figure}

Figure \ref{fig:pre_train_pipeline} illustrates the pre-training pipeline of SDF-TopoNet. In the pre-training phase, the U-Net predicts the SDF representation for each image. The network is divided into a feature extraction backbone and a segmentation head hat outputs the SDF. The predicted SDF is then compared with the ground truth SDF using an MSE loss, allowing the model to learn global topological features and local details efficiently.

\begin{figure}[!ht]
\centering
\resizebox{0.85\linewidth}{!}{%
\includegraphics[width=1.0\linewidth]{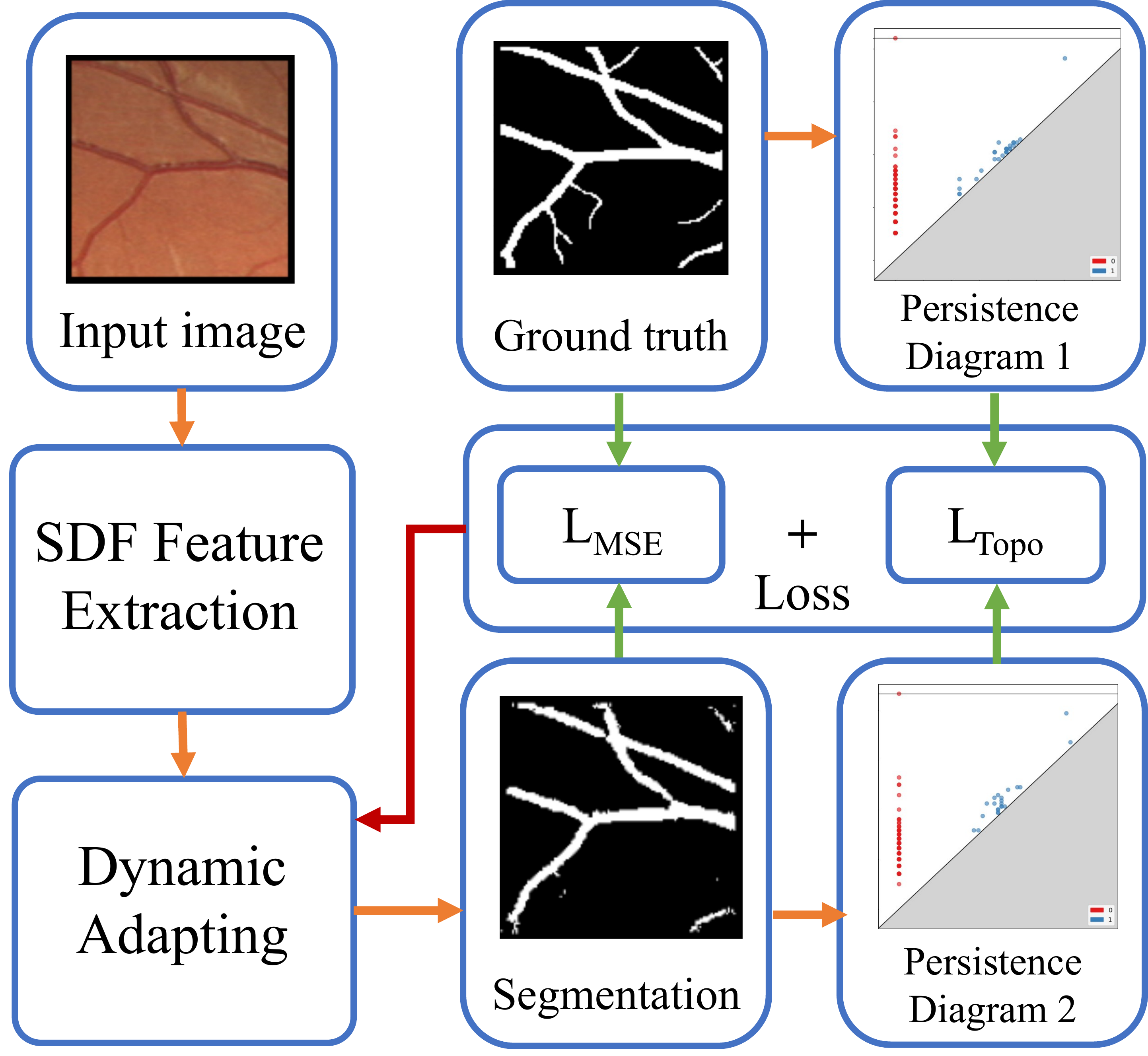}}
\caption{SDF-TopoNet fine-tuning phase. A dynamic adapter transforms SDF features into a segmentation mask. Training optimizes both Dice and topological losses.}
\label{fig:fine_tune_pipeline}
\end{figure}

Figure \ref{fig:fine_tune_pipeline} illustrates the fine-training pipeline of SDF-TopoNet. In the fine-tuning phase, we replace the original segmentation head with a dynamic adapter that maps the learned SDF to a segmentation mask by flattening, applying a fully connected layer, and using a tanh activation. The training objective includes a topological loss \(\,L_{\mathrm{topo}}\), calculated via persistence diagrams, and a Dice loss \(\,L_{\mathrm{Dice}}\). This combination preserves structural consistency while retaining precise boundaries.

\subsection{Signed Distance Function}

The \textbf{signed distance function (SDF)} \cite{malladi1995shape} measures the distance of each point from an object boundary, with the sign indicating whether the point is inside or outside the object:

\begin{equation}
\begin{aligned}
&|f(x)| = g(x) = \inf _{y \in \Omega}\|x-y\|, x \in \mathbb{R}^{\mathrm{n}}\\
&f(x)= 
\begin{cases}
g(x), & \text{if } x \in \Omega, \\ 
-g(x), & \text{if } x \notin \Omega,
\end{cases}
\end{aligned}
\label{equ:sdf_2}
\end{equation}

where \(\Omega\) is the set of points inside the object.

\textbf{Why use SDF as a pre-training target?} A key challenge in topology-aware segmentation is designing an efficient way to preserve topological accuracy. Existing approaches employ topological losses based on persistent homology, but these losses are computationally expensive, making end-to-end training inefficient. Instead of directly applying topological constraints during the entire training process, we introduce a two-stage training strategy. For this approach to be effective, selecting a suitable pre-training target is crucial, and the signed distance function (SDF) naturally emerges as an optimal choice due to its ability to encode topological and geometric information in a smooth and structured manner.

Many topological losses, such as persistence diagram-based loss and Betti matching, rely on distance filtration to extract structural features. Since SDF can encode distance-based topology information and inherently represent the distance to object boundaries, it provides a strong prior that aligns well with topological optimization. Specifically:
\begin{itemize}
\item \textbf{Continuous topological encoding:} Unlike binary segmentation masks, which are discrete and sensitive to local pixel variations, SDF defines a continuous field that captures shape structures more robustly.
\item \textbf{Compatibility with persistent homology:} Since persistent homology methods rely on distance-based filtering, SDF naturally enhances the effectiveness of topological constraints in later training stages.
\item \textbf{Lower early-stage cost:} By pre-training the model to predict SDF, we provide an implicit topological prior that guides structural learning. Without such a prior, training solely with pixel-wise losses like MSE may lead the model to focus excessively on local pixel correctness while neglecting global topological structures. This strategy reduces the need for computationally expensive topological loss in the early stage while still preserving essential topological information.
\end{itemize}

\begin{figure}[!ht]
\centering
\includegraphics[width=1.0\linewidth]{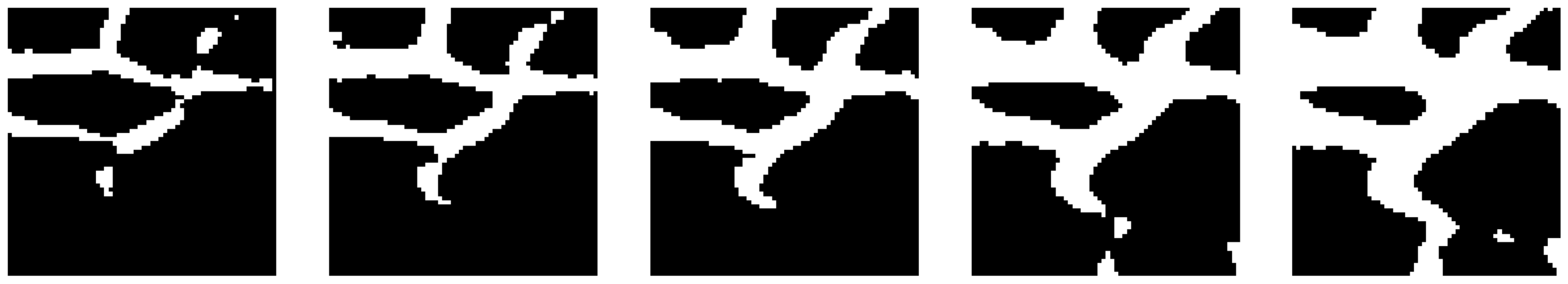}
\caption{Impact of varying SDF thresholds on the segmentation mask. Thresholds from left to right: \(>0.0,>-1.0,>-2.0,>-3.0,>-4.0\). Lowering the threshold progressively enlarges the segmented region, helping to correct broken structures.}
\label{fig:why_sdf_fig1}
\end{figure}

\textbf{Leveraging SDF to refine topology:} As shown in Figure \ref{fig:why_sdf_fig1}, applying different SDF thresholds allows us to manipulate segmentation masks in a topologically meaningful way. Lowering the threshold expands the segmented region, effectively closing gaps in broken structures, which helps reinforce connectivity. This property enables the design of a dynamic adapter in the fine-tuning phase, which transforms SDF values into precise binary masks.

\subsection{Euclidean Distance Transform}

In the pre-training phase of SDF-TopoNet, we prepare the training images by cropping them into smaller patches. To generate the signed distance functions (SDFs) required for training, we compute the Euclidean Distance Transform (EDT) \cite{2020SciPy-NMeth} of the masks. The EDT measures the shortest Euclidean distance from each pixel to the nearest object boundary and is formally defined as:

\begin{equation}
\mathrm{EDT} (x) = \min_{y \in \Omega} \| x - y \|,
\label{equ:edt}
\end{equation}

where $x$ is a pixel in the image, and $\Omega$  represents the set of relevant foreground or background pixels. To obtain the final SDF representation, we compute the EDT separately for both the foreground and background regions and then take their difference:

\begin{equation}
\mathrm{SDF} (x) = \mathrm{EDT}_\mathrm{background} (x) - \mathrm{EDT}_\mathrm{foreground} (x).
\label{equ:sdf_3}
\end{equation}

Here, SDF values are positive inside the object, zero on the boundary, and negative outside, offering a continuous representation of the distance to object boundaries.

\subsection{Dynamic Adapter}

In the pre-training phase, we train a U-Net \cite{ronneberger2015u} model on the preprocessed dataset to predict the signed distance function (SDF) of the target structures. U-Net is widely used for medical image segmentation due to its strong performance on small datasets and efficient training.

In the fine-tuning phase, we replace the original segmentation head with a dynamic adapter, which converts the learned SDF representation into a binary segmentation mask. This step is crucial for integrating topological constraints into the model. 

\textbf{Why a Dynamic Adapter?} 
Since the SDF is a continuous shape representation, the segmentation mask can be obtained by affining SDF values: Manipulating the pre-learned SDF features to assist in fitting the binary ground truth mask. To address this, we introduce a learned channel-wise scaling mechanism that dynamically adjusts based on the SDF features. This mechanism is similar to the Squeeze-and-Excitation module \cite{hu2018squeeze}. 

\begin{figure}[!ht]
\centering
\resizebox{1.05\linewidth}{!}{%
\includegraphics[width=0.85\linewidth]{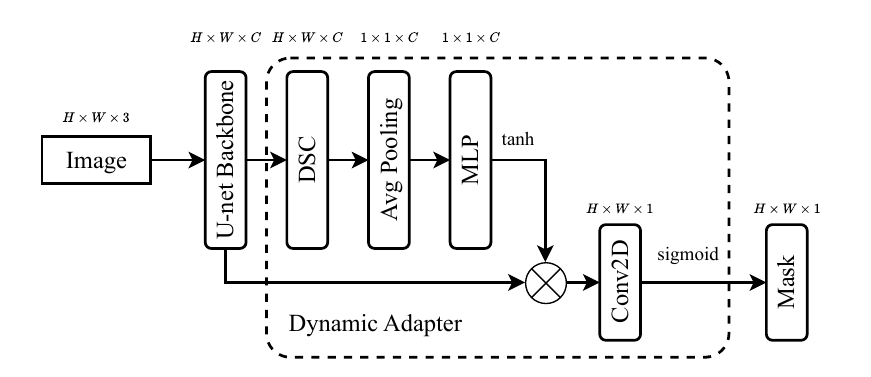}}
\caption{Structure of the dynamic adapter. The feature map undergoes Conv2D-based reconstruction, followed by a fully connected layer that predicts the threshold for segmentation.}
\label{fig:dynamic_threshold}
\end{figure}

As illustrated in Figure~\ref{fig:dynamic_threshold}, the dynamic adapter layer consists of two main components:

\begin{enumerate}
    \item A large-kernel depthwise separable convolution used to learn feature maps obtained from the backbone.
    \item A squeeze module consists of a global average pooling layer and a feed-forward network. It captures information across the entire channel through global pooling and applies channel-wise scaling to the SDF features learned by the backbone.
\end{enumerate}

By replacing a fixed adapter with a learned one, our model adapts to varying structures and improves segmentation quality, particularly in preserving connectivity and reducing fragmentation.

\subsection{Topological Loss}
To ensure topological consistency in segmentation, we employ loss functions based on persistent homology. Given a predicted segmentation mask and its corresponding ground truth, we first compute their persistence diagrams and then measure their topological discrepancy using either the Wasserstein Matching loss or the Betti Matching loss. 

\begin{figure}[!ht]
    \centering
    \resizebox{0.8\linewidth}{!}{
    \includegraphics{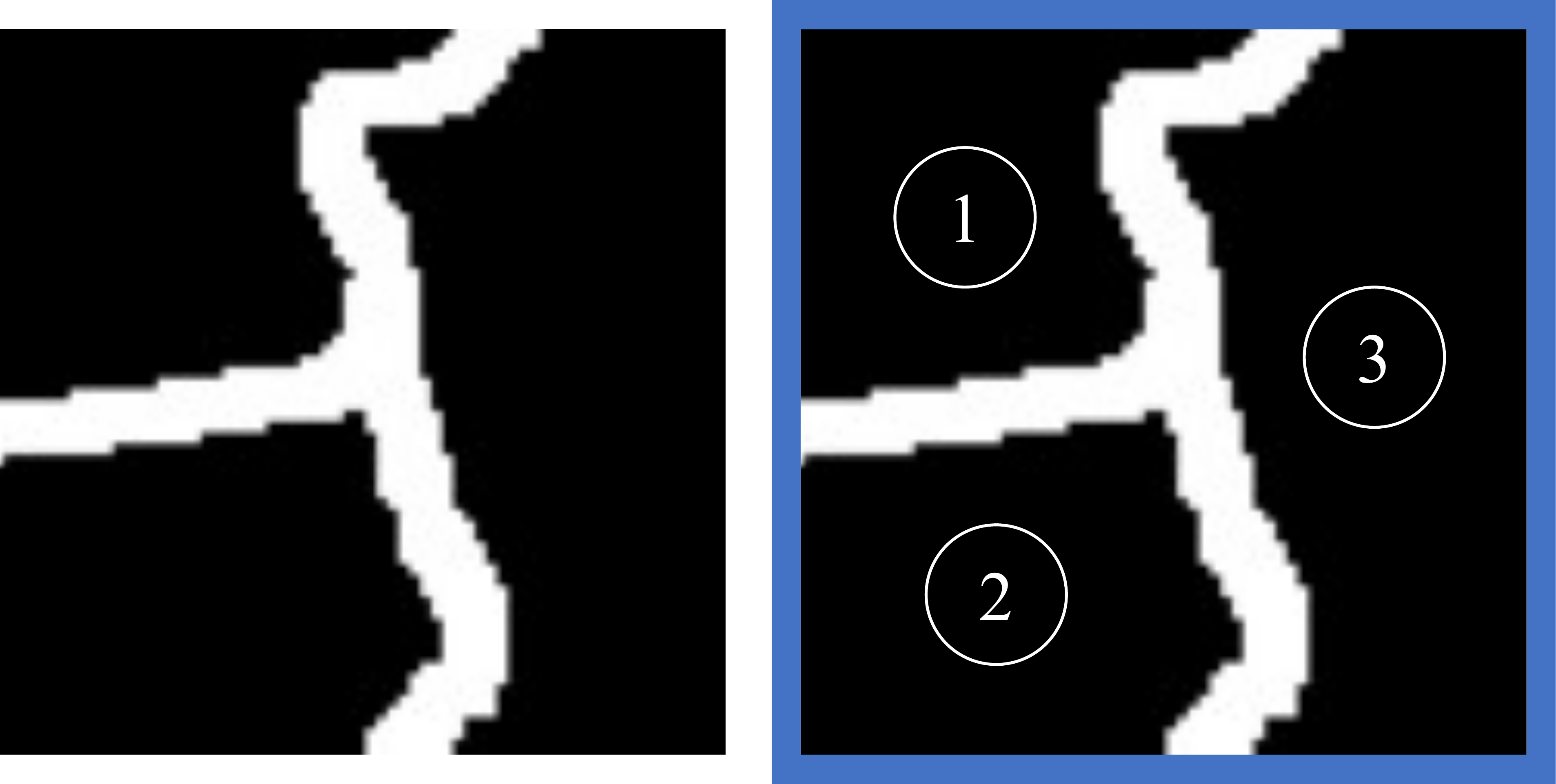}}
    \caption{Padding a bounding box around the mask adds additional topological features, improving the effectiveness of Wasserstein Matching and Betti Matching~\cite{hu2019topology}.}
    \label{fig:padding}
\end{figure}

Since topological loss functions often introduce high computational costs, we evaluate how our proposed training strategy impacts their efficiency and effectiveness in preserving structural accuracy. To stabilize computation, we apply a padding technique as shown in Figure~\ref{fig:padding}, where a $2$-pixel-wide bounding box is added to each mask. This increases the number of points in the persistence diagram, improving the robustness of topological alignment.

\subsubsection{Wasserstein Matching}
Each persistence diagram is represented as a set of points in the birth-death plane. The Wasserstein distance, also known as the Earth Mover’s Distance, measures the discrepancy between two diagrams $D_1$ and $D_2$:

\begin{equation}
W_p\left(D_1, D_2\right) = \left(\inf_{\gamma \in \Gamma\left(D_1, D_2\right)} \sum_{\left(c,c^\prime\right) \in \gamma} \|c - c^\prime\|^p \right)^{1 / p},
\label{equ:wasserstein}
\end{equation}

where $\Gamma(D_1, D_2)$ represents the set of valid point matchings, and $\|\cdot\|$ denotes the Euclidean distance. This metric ensures that the predicted segmentation maintains global topological structures similar to the ground truth.

\subsubsection{Betti Matching}
Betti Matching aligns topological features by establishing correspondences between the Betti numbers of the predicted segmentation and the ground truth. Given a likelihood map $L \in [0,1]^{m \times n}$ predicted by the network, we obtain a binarized segmentation $P \in \{0,1\}^{m \times n}$ and compare it with the ground truth mask $G \in \{0,1\}^{m \times n}$. 

To construct a common filtration space, we define a comparison image $C = \min(P, G)$ (element-wise minimum), ensuring that the induced matchings respect the inclusion relations:

\begin{equation}
\mu(P, G) = \sigma(G, C)^{-1} \circ \sigma(P, C),
\label{equ:betti_matching}
\end{equation}

where $\sigma(P, C): \mathcal{B}(P) \to \mathcal{B}(C)$ and $\sigma(G, C): \mathcal{B}(G) \to \mathcal{B}(C)$ are the induced mappings, and $\sigma(G, C)^{-1}$ is the inverse mapping. The Betti Matching loss is then defined as:

\begin{equation}
L_{\mathrm{BM}}(P, G) = \sum_{q \in \mathrm{Dgm}(P)} 2\|q - \mu(P, G)(q)\|^2.
\label{equ:betti_loss}
\end{equation}

Here, $\mathrm{Dgm}(P)$ represents the persistence diagram of the predicted segmentation, and the factor of 2 normalizes the matching error.

\subsubsection{Topological Loss Function}
We integrate these topological loss functions with Dice loss to balance pixel-wise accuracy and topological consistency:

\begin{equation}
L(s, g) = \alpha \, L_M(s, g) + (1-\alpha) \, L_{\mathrm{Topo}}(s, g),
\label{equ:topo_loss}
\end{equation}

where $s$ is the predicted segmentation, $g$ is the ground truth, $\alpha$ is a weighting factor, and $L_{\mathrm{Topo}}$ is selected as:

\begin{equation}
L_{\mathrm{Topo}}(s, g) =
\begin{cases}
L_{\mathrm{WM}}(s, g), & \text{if Wasserstein Matching}, \\
L_{\mathrm{BM}}(s, g), & \text{if Betti Matching}.
\end{cases}
\end{equation}

By comparing both loss functions, we assess how our two-stage training approach reduces the computational burden while maintaining or improving segmentation accuracy.

\subsection{Fine-tuning with a dynamic adapter and topological loss function}
During fine-tuning, we replace the original U-Net segmentation head with the dynamic adapter. We then fine-tune the entire model using the topological loss. Note that, in this phase, the labels are the ground truth masks rather than the SDFs used in pre-training.

\section{Experiments}

\subsection{Experimental Setup}

We evaluate our method on four datasets: DRIVE \cite{1282003}, CREMI \cite{CREMI2016}, Massachusetts Roads \cite{MnihThesis}, and Elegans \cite{ljosa2012annotated}, covering diverse domains such as biomedical imaging and remote sensing. To assess the effectiveness of SDF as a pre-training prior and the two-stage training strategy, we conduct ablation experiments to analyze their contributions systematically.

\subsubsection{Dataset}

\begin{enumerate} 
\item \textbf{DRIVE:} As shown in \cref{fig:drive}. A retinal vessel segmentation dataset with 40 color fundus images from a diabetic retinopathy screening program. The dataset features intricate vessel structures, making it ideal for evaluating curvilinear pattern segmentation.
\item \textbf{CREMI:} As shown in \cref{fig:cremi_1}. An electron microscopy neuron segmentation dataset with complex, network-like structures and numerous cycles, testing the model’s ability to maintain topological consistency.
\item \textbf{Roads:} A road segmentation dataset with elongated, interconnected networks, similar to CREMI in its topological challenges.
\item \textbf{Elegans:} A biomedical dataset with a balanced mix of dimension-0 and dimension-1 features, useful for evaluating both connected components and topological loops.
\end{enumerate}

\subsection{dataset example}
\begin{figure}[!ht]
    \centering
    \includegraphics[width=1\linewidth]{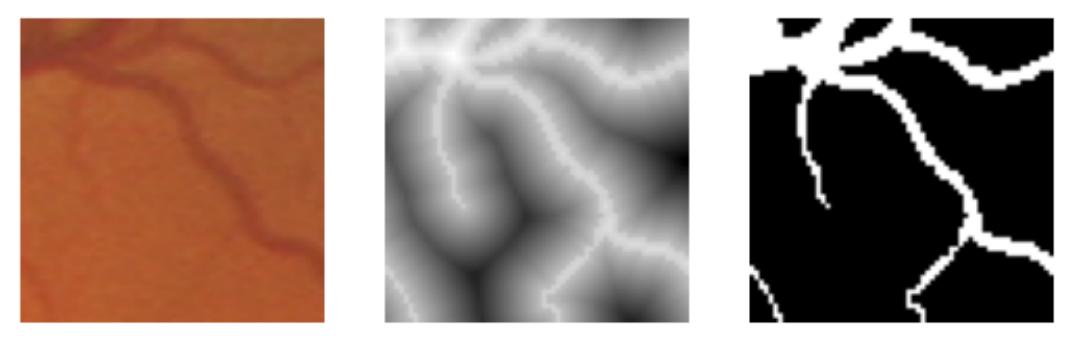}
    \caption{A sample of DRIVE data. From left to right: the original image, the SDF generated from the ground truth, and the ground truth.}
    \label{fig:drive}
\end{figure}

\begin{figure}[!ht]
    \centering
    \includegraphics[width=1.0\linewidth]{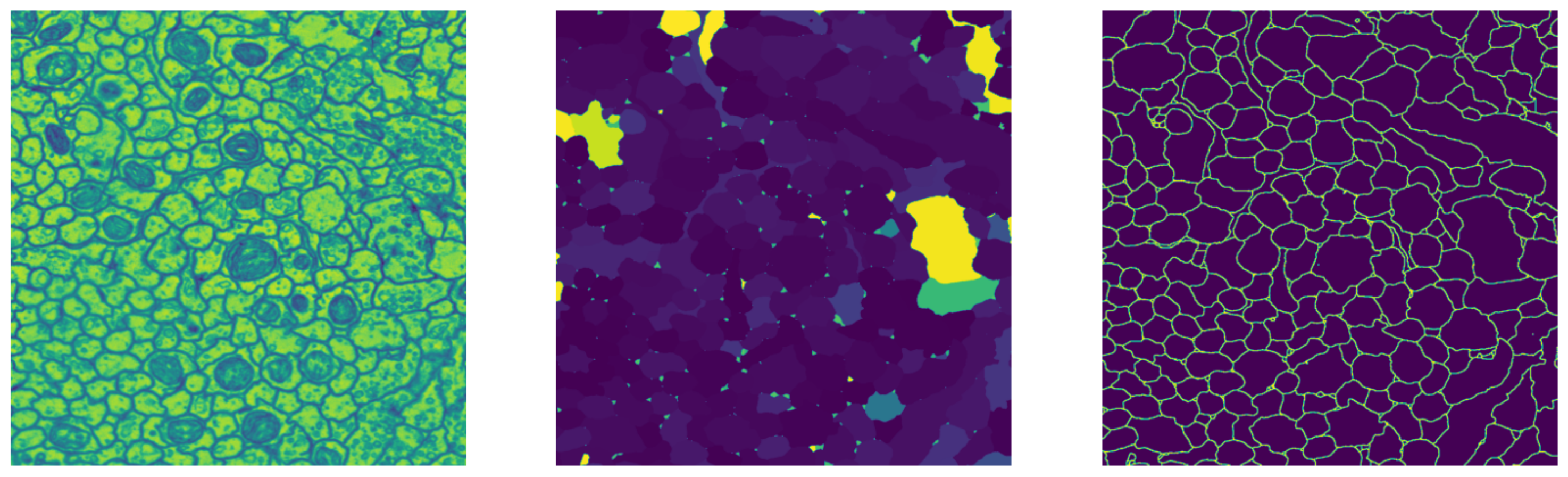}
    \caption{An example of ground truth generated from the original annotations. From left to right: the original slice image, the original annotation, and the ground truth generated from the original annotation.}
    \label{fig:cremi_1}
\end{figure}

We illustrate examples of the two kinds of datasets in Figures~\ref{fig:drive}, \ref{fig:cremi_1}. Specifically, Figure~\ref{fig:drive} shows the DRIVE dataset, while Figures~\ref{fig:cremi_1} depict the CREMI dataset.

\subsubsection{Evaluation Metrics}
We evaluate the models using IoU, PA, and clDice \cite{shit2021cldice} metrics. IoU and PA are commonly used to evaluate semantic segmentation results. clDice and betti matching error, on the other hand, is a good evaluator of topological accuracy. 

\begin{enumerate}
    \item \textbf{IoU (Intersection over Union):} Measures segmentation accuracy by computing the intersection over the union of predicted and true masks.

    \item \textbf{PA (Pixel Accuracy):} Evaluates how many pixels are correctly classified relative to the total number of pixels.

    \item \textbf{Dice (Dice Similarity Coefficient, DSC):} Measures the overlap between the predicted and ground truth masks, balancing precision and recall. 
    
    \item \textbf{clDice:} Used to evaluate the topology score of the prediction mask \cite{shit2021cldice}. To compute clDice, topology precision, and topology sensitivity are proposed. Let $V_L$ be the ground truth, $V_P$ be the predicted result, $S_L$ be the skeleton extracted from the ground truth, and $S_P$ be the skeleton extracted from the predicted result.

    \item \textbf{VoI (Variation of Information):} Measures the distance between two segmentations by quantifying the information loss and gain when transitioning from one to another. Given the entropy of ground truth segmentation $H(V_L)$ and predicted segmentation $H(V_P)$, and their mutual information $I(V_L, V_P)$. 
    
\end{enumerate}

\subsubsection{Benchmark}

We compare \textbf{SDF-TopoNet} against state-of-the-art topology-aware and traditional segmentation baselines:

\begin{enumerate} 
    \item \textbf{Dice:} A standard U-Net model trained with mean Dice similarity coefficient (Dice) loss, serving as the simplest segmentation baseline.

    \item \textbf{Wasserstein Matching:} Extends the U-Net baseline by introducing Wasserstein Matching Loss (WM) for improved topological consistency.

    \item \textbf{Betti Matching:} Augments the U-Net baseline with Betti Matching Loss (BM) to better represent global structural features.
\end{enumerate}

Table~\ref{tab:models} provides a detailed overview of the models and training configurations used in our experiments. The Model column specifies the neural network architecture, where DT represents the Dynamic Adapter. The SDF column indicates whether the model was pre-trained using signed distance functions. The Loss column lists the applied loss functions, with WM denoting Wasserstein Matching and BM referring to Betti Matching. The Alpha column gives the weighting factor ($\alpha$) in the combined loss function, where $\alpha = 1.0$ indicates that only MSE or Dice is used. Our proposed SDF-TopoNet follows a two-phase training paradigm: pre-training with MSE on SDFs and fine-tuning with topology-aware losses while integrating the dynamic thresholding layer.

\begin{table*}[!ht]
\centering
\resizebox{0.8\textwidth}{!}{ 
\fontsize{6pt}{8pt}\selectfont
\begin{tabular}{cccccc}
\toprule
\multicolumn{2}{c}{\textbf{Method}} & \textbf{Model} & \textbf{SDF} & \textbf{Loss} & \textbf{Alpha} \\ 
\hline
\multicolumn{2}{c}{U-net}    & U-net & disable & Dice        & 1.0 \\ 
\hline
\multicolumn{2}{c}{Wasserstein Matching} & U-net & disable  & Dice + WM & 0.9 \\
\hline
\multicolumn{2}{c}{Betti Matching} & U-net & disable  & Dice + BM & 0.9 \\
\hline
\multirow{2}*{Ours (WM)} & Pre-train   & U-net & enable  & MSE        & 1.0\\
\cline{2-6}
                          & Fine-tune   & U-net + DT & disable  & Dice + WM  & 0.9\\
\hline
\multirow{2}*{Ours (BM)} & Pre-train   & U-net & enable  & MSE        & 1.0\\
\cline{2-6}
                          & Fine-tune   & U-net + DT & disable  & Dice + BM  & 0.9\\
\bottomrule
\end{tabular}
}
\caption{Comparison of benchmark methods and their configurations. SDF denotes whether the model is pre-trained with a signed distance function (enable/disable). DT refers to the inclusion of a dynamic thresholding layer. The loss entries list the cost functions employed, and \(\alpha\) represents the weight balancing among different loss terms.}
\label{tab:models}
\end{table*}

\subsection{Implementation details}
We split the dataset into training, validation, and test sets with a 14:3:3 ratio. During Wasserstein matching, we randomly sample patches of size 64. For Betti matching, we use patches of size 32 due to higher training costs. At test time, data are sampled with a sliding window approach. The backbone of our U-Net model is EfficientNet-B0, trained on a single NVIDIA A40 GPU with the Adam~\cite{kingma2017adammethodstochasticoptimization} optimizer. We pre-train the model for 200 epochs using mean-squared error (MSE) loss on SDFs, then fine-tune for 100 epochs with a topological loss. This schedule reduces overall computational overhead.

\subsection{Results}
Table~\ref{tab:aStrangeTable} compares our approach against existing baselines. Our method shows strong gains, especially on datasets with prominent connectivity structures. For instance, on the DRIVE dataset, SDF-TopoNet (WM) raises the Dice score from 0.267 (Stucki et al.) to 0.716, a 168\% relative improvement. In the CREMI dataset, our WM variant achieves a Dice of 0.872, surpassing both the Dice baseline (0.790) and Stucki et al. (0.832).

\begin{table*}[!ht]
\centering
\resizebox{0.8\textwidth}{!}{
\fontsize{6pt}{4pt}\selectfont
\begin{tabular}{l|c|ccccc}
\toprule
 Dataset & Method & Dice $\uparrow$ & IoU $\uparrow$ & PA $\uparrow$ & VoI $\downarrow$ & clDice $\uparrow$\\
\midrule
\parbox[t]{6mm}{\multirow{5}{*}{DRIVE}} 
& Dice          & 0.243 & 0.149 & 0.572 & 0.974 & 0.308 \\
& Hu et al.     & 0.123 & 0.079 & 0.539 & \textbf{0.535} & 0.155 \\
& Stucki et al. & 0.267 & 0.179 & 0.596 & 0.840 & 0.326 \\
& Ours (WM)     & \textbf{0.716} & \textbf{0.575} & \textbf{0.847} & 0.539 & \textbf{0.758} \\
& Ours (BM)     & 0.606 & 0.480 & 0.803 & 0.592 & 0.674 \\
\midrule
\parbox[t]{6mm}{\multirow{5}{*}{CREMI}} 
& Dice          & 0.790 & 0.668 & 0.850 & 0.558 & 0.916 \\
& Hu et al.     & 0.849 & 0.742 & 0.884 & 0.585 & 0.877 \\
& Stucki et al. & 0.832 & 0.719 & 0.912 & 0.590 & 0.891 \\
& Ours (WM)     & \textbf{0.872} & \textbf{0.775} & 0.930 & \textbf{0.498} & \textbf{0.929} \\
& Ours (BM)     & 0.855 & 0.751 & \textbf{0.935} & 0.543 & 0.919 \\
\midrule
\parbox[t]{6mm}{\multirow{5}{*}{Roads}}
& Dice          & 0.598 & 0.453 & 0.818 & \textbf{0.406} & 0.659 \\
& Hu et al.     & 0.582 & 0.411 & 0.815 & 0.532 & 0.636 \\
& Stucki et al. & 0.537 & 0.395 & 0.828 & 0.608 & 0.605 \\
& Ours (WM)     & \textbf{0.616} & \textbf{0.469} & \textbf{0.841} & 0.413 & \textbf{0.690} \\
& Ours (BM)     & 0.574 & 0.433 & 0.839 & 0.559 & 0.644 \\
\midrule
\parbox[t]{6mm}{\multirow{5}{*}{Elegans}}
& Dice          & 0.863 & 0.762 & 0.901 & 0.282 & 0.939 \\
& Hu et al.     & 0.817 & 0.710 & 0.872 & 0.310 & 0.901 \\
& Stucki et al. & 0.838 & 0.735 & 0.889 & 0.305 & 0.910 \\
& Ours (WM)     & \textbf{0.880} & \textbf{0.786} & \textbf{0.933} & \textbf{0.282} & \textbf{0.951} \\
& Ours (BM)     & 0.867 & 0.777 & 0.929 & 0.290 & 0.935 \\
\bottomrule
\end{tabular}
}
\caption{Quantitative results for each model on four datasets, evaluating Dice, IoU, PA, VoI, and clDice metrics, where WM and BM respectively represent the use of Wasserstein Matching and Betti Matching as the topology loss in the fine-tuning stage.}
\label{tab:aStrangeTable}
\end{table*}

\begin{table*}[!ht]
\centering
\resizebox{0.75\textwidth}{!}{
\fontsize{6pt}{4pt}\selectfont
\begin{tabular}{l|ccc}
\toprule
  Method & Epochs & Runtime/epoch & Total time \\
\midrule
Dice          & 100 & 1.2s &   1m57s\\
Hu et al.     & 100 & 11.2s &   18m37s\\
Stucki et al. & 100 & 1m27.8s  &  2h26m19s\\
Ours (WM)     & 200P1 + 100P2 & 1.3s + 11.2s  &   4m12s + 18m37s \\
Ours (BM)     & 200P1 + 100P2 & 1.3s + 1m27.8s &  4m12s + 2h26m19s \\
\bottomrule
\end{tabular}
}
\caption{The running time performance of each method. Our method divides the two training stages with a “+”.}
\label{tab:runtime}
\end{table*}

A key benefit of our two-stage approach is its ability to substantially boost segmentation performance with only a small increase in training time. As shown in Table~\ref{tab:runtime}, topological losses are generally expensive to compute. For example, Stucki et al. (Betti Matching) requires 2h26m19s, whereas our Betti Matching model (SDF-TopoNet BM) attains better results in 2h30m31s—an additional 4m12s for pre-training that pays off in both pixel-level and topological accuracy. Similarly, for Wasserstein Matching, Hu et al.’s model finishes in 18m37s, while our SDF-TopoNet WM model outperforms it in 22m49s, again adding only 4m12s in pre-training.

Finally, we find that SDF pre-training may speed up convergence during the fine-tuning phase. Because we already surpass the competing methods with the current schedule, we expect SDF-TopoNet could deliver similar or better results with fewer fine-tuning epochs. This indicates the potential for further optimizing training efficiency without harming segmentation performance.

\subsection{Ablation study}

To validate the effectiveness of our method, we conduct an ablation study on the CREMI dataset by evaluating the impact of using SDF as a pre-training target. In this experiment, we simply split the training into two stages while keeping the pre-training target identical to the final segmentation mask. This allows us to isolate the contribution of SDF pre-training in capturing topological structures.

The results, presented in Table~\ref{tab:ablation_sdf}, indicate that incorporating SDF pre-training improves both topological and pixel-wise accuracy. Specifically, Dice and IoU scores increase from 0.842 to 0.872 and from 0.734 to 0.775, respectively, demonstrating improved segmentation quality. Additionally, the clDice metric, which measures topological consistency, also improves from 0.897 to 0.929. Meanwhile, Variation of Information (VoI) decreases, suggesting a reduction in segmentation uncertainty.

\begin{table}[!ht]
\centering
\resizebox{1.0\columnwidth}{!}{
\fontsize{6pt}{4pt}\selectfont
\begin{tabular}{l|c|ccccc}
\toprule
 Loss & SDF & Dice $\uparrow$ & IoU $\uparrow$ & PA $\downarrow$ & VoI $\uparrow$ & clDice $\uparrow$\\
\midrule
\multirow{2}{*}{Hu et al.}
& Enable      & \textbf{0.872} & \textbf{0.775} & \textbf{0.930} & \textbf{0.498} & \textbf{0.929} \\
& Disable     & 0.842 & 0.734 & 0.929 & 0.568 & 0.897 \\
\bottomrule
\end{tabular}
}
\caption{Ablation study on the effect of enabling or disabling SDF for the methods by Hu et al. and Stucki et al. We report Dice, IoU, Pixel Accuracy (PA), Variation of Information (VoI), and clDice.}
\label{tab:ablation_sdf}
\end{table}

These findings suggest that SDF pre-training provides a meaningful intermediate representation that aids in learning structural features more effectively. By first learning an SDF representation, the model can better capture topological structures, leading to more accurate and consistent segmentations. This further supports our hypothesis that SDF serves as a useful auxiliary target that balances topological correctness and computational efficiency.

\section{Conclusion}
We propose SDF-TopoNet, a two-stage framework that merges SDF pre-training with topology-aware fine-tuning to improve segmentation accuracy and manage computational costs. By pre-training on the signed distance function using a simple MSE loss, the network learns essential geometric and boundary features before applying topological constraints, thus avoiding expensive persistent homology operations in early stages. This approach addresses two major challenges in topology-aware segmentation: it reduces the computational load by limiting topological losses to the fine-tuning phase, and it preserves both pixel-level detail and high-level structural integrity. Experiments on the DRIVE, CREMI, Roads, and Elegans datasets confirm that SDF-TopoNet delivers enhanced topological performance without sacrificing pixel accuracy. Overall, our results highlight the potential of SDF-based pre-training as a scalable strategy for integrating topological information into segmentation models.

\clearpage
\bibliographystyle{unsrt}
\bibliography{ref}

\clearpage

\renewcommand\thesection{A\arabic{section}}
\renewcommand\thefigure{A\arabic{figure}}
\renewcommand\thetable{A\arabic{table}}  
\renewcommand\theequation{A\arabic{equation}}
\setcounter{section}{0}
\setcounter{equation}{0}
\setcounter{table}{0}
\setcounter{figure}{0}

\setcounter{page}{1}
\maketitlesupplementary

\section{Background}\label{appendix:background}

Before introducing our method, we provide a brief overview of the topological data analysis (TDA) concepts that are essential for understanding our approach~\cite{chazal2021introductiontopologicaldataanalysis}. We begin by discussing homology and Betti numbers, then describe persistent homology and its application to computing persistence diagrams.

\subsection{Homology and Betti Numbers}
Homology is a fundamental tool in algebraic topology that captures the connectivity and shape of a space by associating algebraic structures to it. Specifically, homology assigns a sequence of abelian groups or vector spaces to a topological space, describing its global structure in different dimensions.

For a given cubical complex \( K \), which is well-suited for image analysis, we define its chain complex \( C_*(K) \) as follows: 

\begin{equation}
C_*(K) = \bigl(\{C_d(K)\}_{d \in \mathbb{Z}}, \{\partial_d\}_{d \in \mathbb{Z}}\bigr),
\end{equation}
where \( K_d \) is the set of \( d \)-dimensional cells in \( K \), and let \( C_d(K) \) be the vector space over \( \mathbb{Z}_2 \) generated by \( K_d \). The boundary operator \( \partial_d: C_d(K) \to C_{d-1}(K) \) maps each \( d \)-cell to a sum of its \((d-1)\)-dimensional faces. The boundary operators satisfy the fundamental following property: 
\begin{equation}
\partial_{k-1} \circ \partial_k \equiv 0 \ \ \mbox{for any} \ \  k \geq 1. 
\end{equation}
We then define the cycle and boundary subspaces as:

\begin{equation}
Z_d(K) = \mathrm{Ker}(\partial_d), \quad B_d(K) = \mathrm{Im}( \partial_{d+1}),
\end{equation}
where the $\mathrm{Ker}$ and $\mathrm{Im}$ is the kernel and image of the operator respectively. This leads to the following definition:

\begin{definition}[Homology group and Betti numbers]
The $k^{th}$ (simplicial) \textbf{Homology group} of $K$ is the quotient vector space
\begin{equation}
H_k(K) = Z_k(K) / B_k(K). 
\end{equation}
The $k^{th}$ \textbf{Betti number} of $K$ is the dimension $\beta_k(K) = \dim H_k(K)$ of the vector space $H_k(K)$. 
\end{definition}

Simplicial homology groups and Betti numbers are topological invariants: if $K$ and $K^\prime$ are two simplicial complexes whose geometric realizations are homotopy equivalent, their homology groups are isomorphic and their Betti numbers match. In particular: \( \beta_0 \) counts connected components, \( \beta_1 \) counts loops or cycles and \( \beta_2 \) (in 3D) counts voids or enclosed cavities.

\subsection{Persistent Homology}
Persistent homology generalizes classical homology by examining how topological features evolve over multiple scales. Instead of analyzing a single complex, we study a \textbf{filtration}, a nested sequence of complexes:

\begin{equation}
K_0 \subseteq K_1 \subseteq \dots \subseteq K_n.
\end{equation}

As we move through the filtration, topological features (e.g., connected components and loops) appear and disappear. Persistent homology records the birth and death of these features, assigning each one a lifespan interval \([b, d]\), where \( b \) and \( d \) represent the birth and death times, respectively.

Formally, for a function \( f: K \to \mathbb{R} \) that defines a filtration, we construct subcomplexes:

\begin{equation}
D(f)_r = \{c \in K \mid f(c) \leq r\}.
\end{equation}

Each inclusion \( D(f)_r \subseteq D(f)_s \) (for \( r \leq s \)) induces maps on homology groups:

\begin{equation}
H_d(D(f)_r) \to H_d(D(f)_s).
\end{equation}

These maps form a \textbf{persistence module}, which tracks the duration of each homology class across the filtration.

\subsection{Persistence Diagram}
A \textbf{persistence diagram} visually represents the birth and death times of topological features in a filtration. Each feature is plotted as a point \( (b, d) \), where the horizontal coordinate represents the birth time and the vertical coordinate represents the death time.

\begin{figure}[!ht]
    \centering
    \resizebox{0.9\columnwidth}{!}{
    \includegraphics[width=1\linewidth]{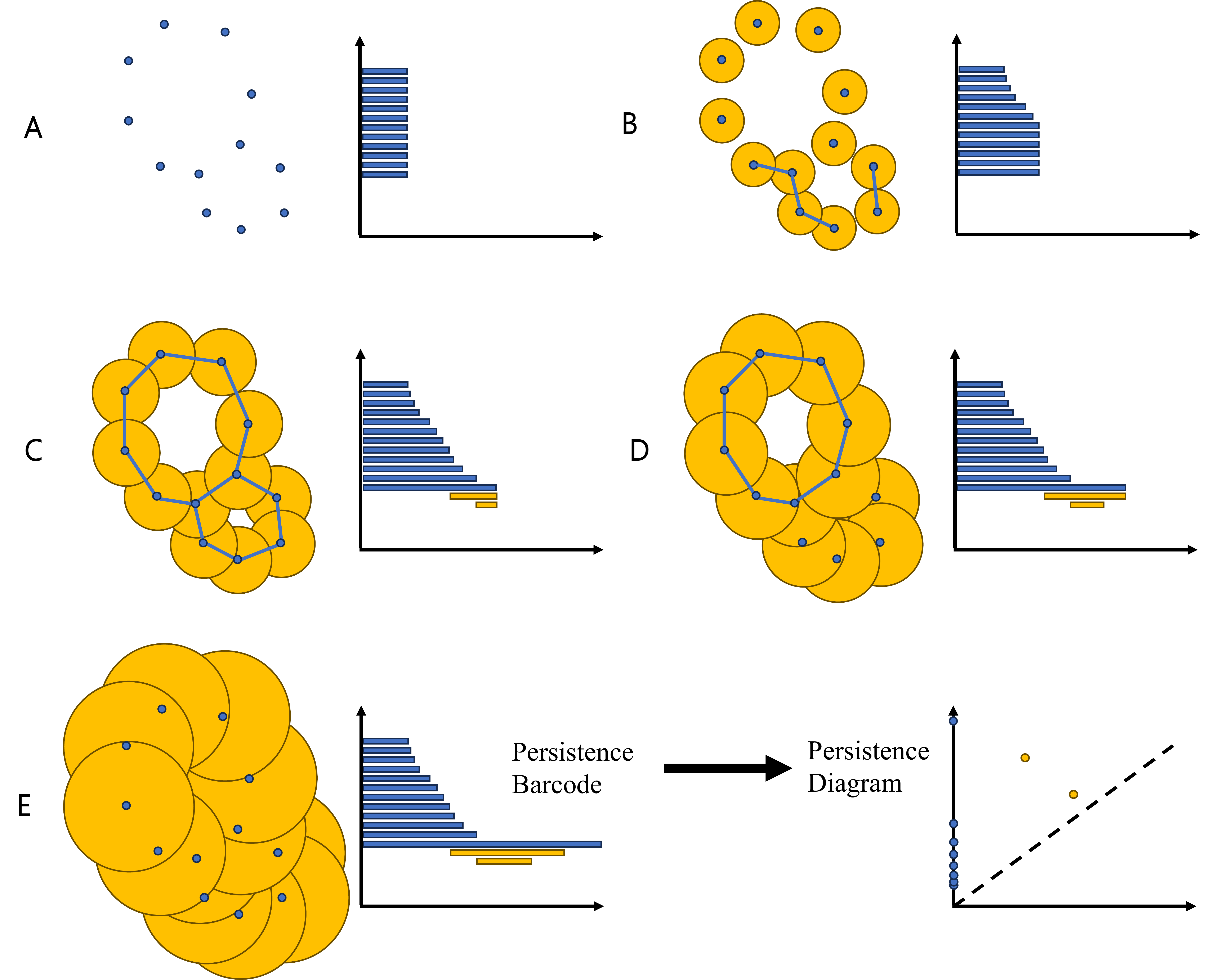}}
    \caption{Generation of a persistence diagram for a simple point cloud. As filtration progresses, topological features emerge and disappear, forming a persistence barcode that can be represented as a persistence diagram.}
    \label{fig:persistence_diagrams}
\end{figure}

Long-lived features (far from the diagonal) are considered significant, while short-lived ones are often regarded as noise.

\subsection{Cubical Complex Representation}
For image-based tasks, it is more natural to use cubical complexes rather than simplicial complexes. A cubical complex treats each pixel as a cell, allowing image structures to be analyzed through topological filtration.

\begin{figure}[!ht]
\centering
\includegraphics[width=1\linewidth]{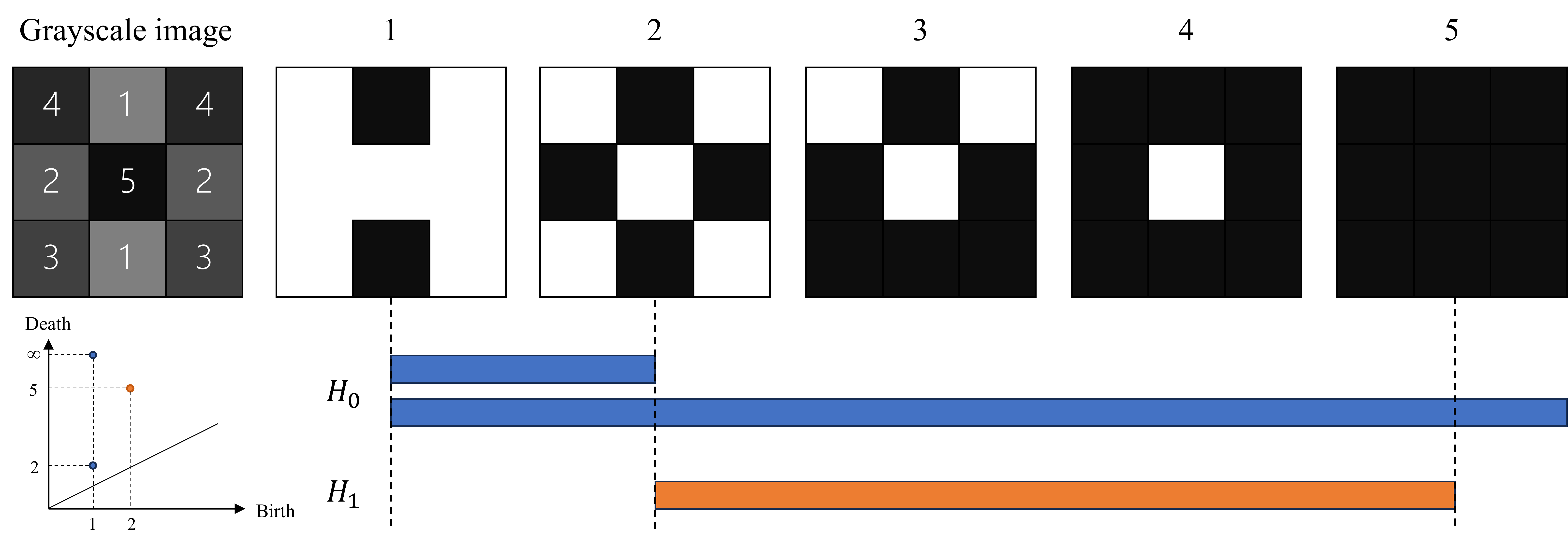}
\caption{Filtration process of a grayscale image using a cubical complex.}
\label{fig:cubical_complex}
\end{figure}

Figure~\ref{fig:cubical_complex} illustrates a cubical filtration applied to a simple $3 \times 3$ grayscale image. As the grayscale threshold increases, the topology of the binary representation evolves, generating a sequence of filtered images.

At a threshold of 1, two separate connected components appear, corresponding to two $H_0$ (blue) bars in the barcode diagram. As the threshold increases to 2, the central four central pixels merge into a single connected component, causing one $H_0$ bar to terminate.At this point, a hole ($H_1$ feature) forms in the center, represented by an orange bar in the barcode. Once the threshold reaches 5, all pixels have joined into one connected component, and the previously formed hole disappears, ending the $H_1$ feature. The persistence diagram summarizes these topological changes by recording the birth and death times of each component and loop.

\end{document}